# Exploiting Queue Information for Quality of Services Routing in SDN Networks

Hong-Li Lin
*Dept. of Electrical Engineering*
*National Taipei University of Technology*
Taipei, Taiwan
t112318154@ntut.edu.tw

Yung-Chung Wang
*Dept. of Electrical Engineering*
*National Taipei University of Technology*
Taipei, Taiwan
ycwang@ntut.edu.tw

**Abstract—Routing costs determine how deterministic networks allocate feasible queue resources to successive flow requests. We propose a topology-aware cost that combines normalized network-calculus post-admission delay and backlog bounds with static path diversity and shortest-path load concentration. Their product weights the backlog premium above a delay baseline within a shared queue-level routing, LARAC, and admission-control framework. Across ten paired runs, the proposed cost admits 657.7, 478.9, and 168.2 more flows than Max-Risk on Polska, NSFNET, and COST266, respectively; the corresponding 95% confidence intervals exclude zero. Germany50 shows a smaller gain of 67.6 flows with an interval crossing zero. The proposed cost has the highest mean acceptance among five costs on three of the four topologies. Gains are accompanied by comparable or lower mean hop counts and cannot be explained simply by higher aggregate utilization. On 100–1000-node networks, risk-aware costs remain comparatively stable while delay-only acceptance declines. A centralized SDN realization reduces distributed control and protocol-state overhead, while large-scale results expose the associated routing-performance tradeoff.**

## I. INTRODUCTION

Deterministic networking supports applications with bounded end-to-end latency through traffic control, resource provisioning, and admission control [1], [2]. Queue-level routing makes the selected priority queue part of a route. Preconfigured queueing-delay budgets provide an additive path-delay metric, while network calculus (NC) checks whether admitting a flow preserves queue delay and buffer constraints [3], [4].

Even with a common feasibility model, a routing decision affects the resources left for later requests. Minimizing post-admission delay alone can overlook backlog pressure. Conversely, minimizing backlog alone can overlook delay pressure. A maximum of normalized delay and backlog captures the dominant resource indicator, but applies the same tradeoff to topologies with different opportunities and needs for rerouting.

This paper asks how static topology structure should condition that tradeoff. Path diversity represents the structural availability of alternative routes; shortest-path load concentration represents the tendency of baseline routing to concentrate demand. We combine these descriptors through a bilinear interaction that scales the backlog premium above a delay-oriented baseline. The cost remains between the normalized delay indicator and the maximum of the two resource indicators.

We make three contributions: a progression from single-resource costs to a joint post-admission delay/backlog risk cost; a topology-conditioned cost using static path diversity P and shortest-path load concentration L through the bilinear interaction $\gamma = P_N L_N$; and an evaluation across 14 topologies covering small-topology performance, P/L ablation, mechanism analysis, and 100–1000-node scalability. Queue-level topology, strict priority queuing (SPQ), NC feasibility, and LARAC [5] remain shared. The centralized SDN realization supports the cost design and a separate architecture comparison.

## II. RELATED WORK

DetNet specifies an architecture and bounded-latency framework for provisioned traffic [1], [2]. NC supplies analytical delay and backlog bounds [4]. Miserez et al. [3] exploit queue information and queue-level delay budgets for scalable delay-constrained routing. DetServ [6] develops deterministic network models for real-time QoS provisioning in SDN, and Chameleon [7] combines queue awareness with adaptive source routing. These works establish that queue state and routing decisions must be considered together.

Recent work addresses complementary optimization problems. DetWAN [8] studies deadline-aware routing and scheduling in wide-area networks. Pulse+ [9] considers DetNet routing under delay-difference constraints. Differentiable NC supports network synthesis under delay constraints [10], while NC-based deterministic routing has also been investigated for LEO satellite networks [11]. Existing work addresses deterministic feasibility, queue-aware routing, and routing/scheduling optimization; this paper isolates routing-cost design under a common queue-level feasibility framework.

## III. SYSTEM MODEL AND ARCHITECTURE

### A. Queue-Level Routing and Feasibility

Let G = (V, A) be the directed data-plane graph, with two arcs for each bidirectional physical link. Each output port has multiple non-preemptive SPQ queues. Expanding a physical arc into its queue alternatives yields a queue-level graph; a path $\pi$ selects forwarding arcs and priorities. Queue q has per-queue DNC buffer-size budget $B_q$ and queueing-delay budget $D_q$, while physical arc e has propagation delay $d_e$. A request specifies source, destination, deadline $D_f$, sustainable rate $r_f$, and burst $b_f$, with arrival envelope $\alpha_f(t) = r_f t + b_f$.

$$d(\pi) = \sum_{q \in \pi} (\tau_q + D_q) \le D_f \quad (1)$$

The service model includes interference from higher priorities and blocking by an already transmitting lower-priority packet. Per-hop traffic envelopes and SPQ service curves follow the shared NC framework [3], [4]. For a tentative admission, let ${\alpha_q}^+$ and ${\beta_q}^+$ denote the aggregate arrival and service curves. The horizontal and vertical deviations give the post-admission queueing-delay bound ${h_q}^+$ and backlog bound ${v_q}^+$:

$$ {h_q}^+ = sup_{t\geq 0}\, inf\{u \geq 0 : {\alpha_q}^+(t) \leq {\beta_q}^+(t+u)\} \tag{2} $$

$$ {v_q}^+ = sup_{t\geq 0}\, \{{\alpha_q}^+(t) - {\beta_q}^+(t)\} \tag{3} $$

Admission checks the selected queue and every affected lower-priority queue on its output port. With $S_q$ denoting this set, the inherited strict constraints are

$$ {h_j}^+ < D_j,\ {v_j}^+ < B_j,\ j \in S_q \tag{4} $$

Nonfinite bounds fail admission. Together, (1)–(4) define the shared deterministic feasibility conditions for new and existing reservations. The cost ranks admissible alternatives; it does not replace those conditions.

### B. *Centralized Workflow*

Fig. 1 presents the conceptual architecture. Routers send periodic and event-triggered Queue-Level State Advertisements (QLSAs), and the controller stores the latest received reports in its Queue-Level State Database (QLSDB). NC post-admission analysis derives queue-state indicators, while static topology supplies P and L. The resulting routing cost enters LARAC with admission control for path and queue assignment.

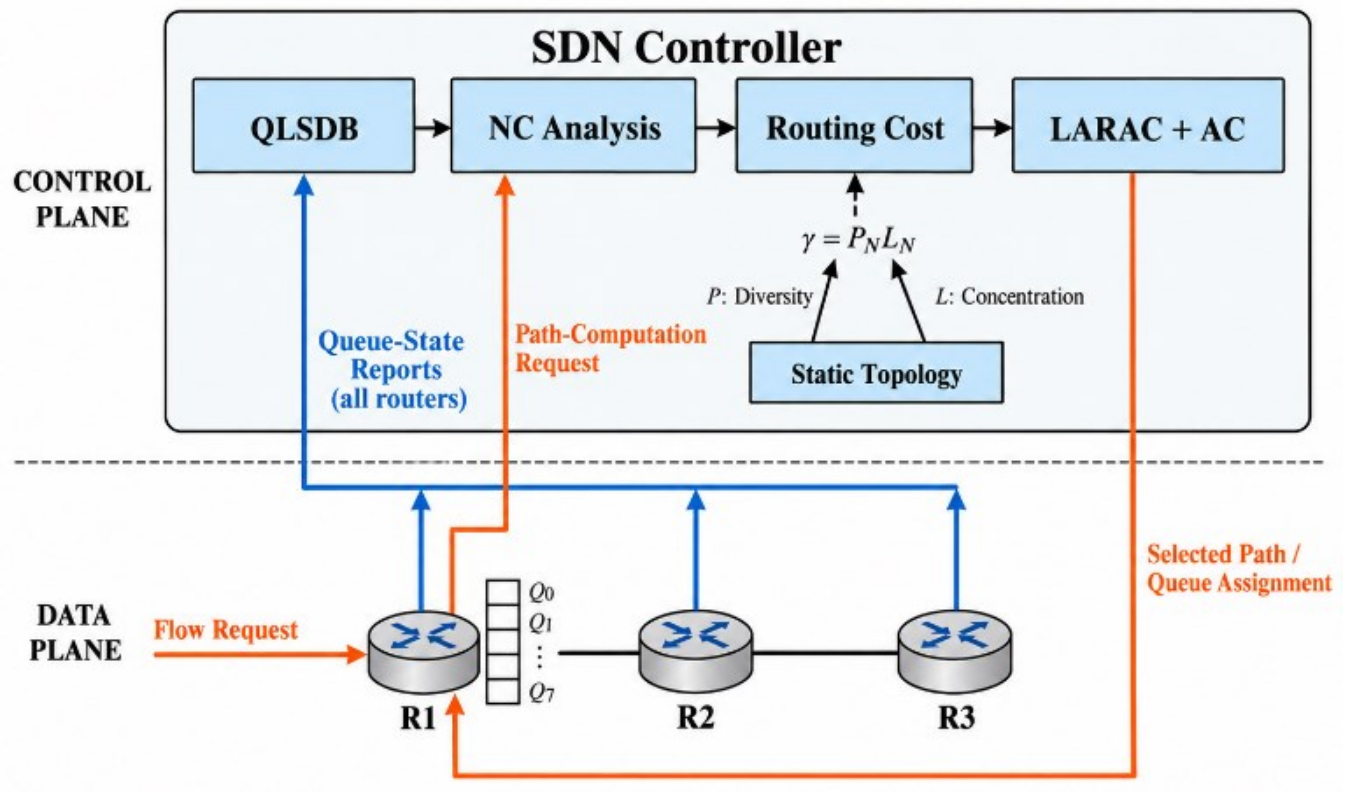

Fig. 1. Logically centralized routing architecture.

For a fixed request and state snapshot, the path objective is the sum of queue-edge costs. LARAC first tests the minimum-cost path and then, when needed, the minimum-delay path. It iteratively searches with an aggregated cost-plus-delay metric to obtain a low-cost delay-feasible candidate [5]. NC-inadmissible queue choices are excluded under the common admission procedure.

QLSDB reports can lag local reservations, so nodes revalidate admission during installation. Failure rolls back partial reservations, refreshes state, and permits retry under the common limit; otherwise the request is rejected. Successful validation embeds the path. All five costs share this procedure.

### C. *Separating Architecture and Cost Effects*

Link-State Dijkstra (LS-Dijkstra) is the distributed baseline: each router maintains disseminated queue/link state and independently runs Dijkstra, following the link-state-style architecture of [3]. SDN-Dijkstra retains the same Dijkstra routing logic but moves maintenance of the network-wide state view and path computation to a logically centralized controller. This comparison isolates distributed versus centralized control. Subsequent experiments then fix centralized control, adopt LARAC with admission control, and compare the five routing costs to isolate cost-design effects.

## IV. TOPOLOGY-AWARE ROUTING COST

In the Dijkstra baselines, queue-level delay serves as the additive shortest-path metric. The subsequent cost experiments use LARAC, which separates the routing objective from the end-to-end delay constraint: it searches for a low-cost path subject to the delay bound through iterative Lagrangian combinations of cost and delay [5]. Within this fixed LARAC and admission-control framework, the following five costs define alternative routing objectives.

### A. *Queue-Risk Baselines*

For each tentative queue assignment, normalize the NC post-admission bounds as

$$ H_q = \frac{{h_q}^+}{D_{ref}},\quad V_q = \frac{{v_q}^+}{B_q} \tag{5} $$

$D_{\text{ref}}$ = 32 ms = 0.032 s, the largest queueing-delay budget in the inherited eight-queue Base configuration [3], is the common delay-normalization scale. A common reference rather than the queue-specific budget keeps identical absolute post-admission delays comparable across priorities; queue-specific normalization would rescale the same delay and weaken priority differentiation. $B_q$ = 97,000 bytes = 776,000 bits is the inherited per-queue DNC buffer-size budget [3]. Analytical indicators $H_q$ and $V_q$ normalize delay and backlog, respectively.

$$ M_q = max(H_q, V_q),\quad \Delta_q = max(0, V_q - H_q) \tag{6} $$

**Delay Cost (Delay):** normalized post-admission delay.

$$ C_D = H \tag{7} $$

**Backlog Cost (Backlog):** normalized post-admission backlog.

$$ C_B = V \tag{8} $$

Both are single-resource views. **Max-Risk Cost (Max-Risk):** the dominant normalized delay/backlog indicator.

$$ C_M = max(H, V) = M \tag{9} $$

### B. *Static Topology Descriptors*

P averages pairwise edge connectivity over all unordered router pairs in the undirected physical data-plane graph $G_u$ =

$(V, E_u)$, with one edge per bidirectional link and no controller nodes or control-plane links:

$$P = \frac{1}{\binom{|V|}{2}} \sum_{\{s,t\} \subset V} \lambda_{G_u}(s,t) \quad (10)$$

Here $\lambda_{G_u}$(s,t) counts the maximum edge-disjoint s–t paths. Higher P means greater rerouting opportunity.

For ordered router pairs s ≠ t, π(s,t) minimizes physical/NED distance. For reproducibility, equal-distance shortest paths are resolved deterministically using natural router-index ordering. Directed arc e ∈ A has load

$$\ell_e = \sum_{s \neq t} 1\{e \in \pi(s,t)\} \quad (11)$$

With 1{·} denoting the indicator, define load concentration over all directed arcs, including zero-load arcs:

$$L = \frac{\sigma_{\text{pop}}(\{\ell_e\}_{e \in A})}{\mu(\{\ell_e\}_{e \in A})} \quad (12)$$

Here μ and $\sigma_{\text{pop}}$ are the population mean and standard deviation. Higher L means less even shortest-path loading: greater rerouting need. Both descriptors are topology-static, independent of runtime arrivals and current queue occupancy. Max-Risk ignores these distinctions, motivating topology conditioning.

### *C. Normalization and Bilinear Interaction*

Rounding the observed minimum and maximum across the evaluated topology set to one decimal place gives $P_{\min} = 2.5$, $P_{\max} = 5.3$, and $L_{\min} = 0.4$, $L_{\max} = 1.2$. Then

$$P_N = \text{clip}\left(\frac{P - P_{\min}}{P_{\max} - P_{\min}}, 0, 1\right) \quad (13)$$

$$L_N = \text{clip}\left(\frac{L - L_{\min}}{L_{\max} - L_{\min}}, 0, 1\right) \quad (14)$$

Here clip(x, 0, 1) = min(1, max(0, x)). Values outside the rounded intervals are clipped to [0, 1]. Zero denotes the lower range, not absent alternatives or concentration.

**Diversity Cost (P-only):** weights Δ from the Delay baseline by rerouting opportunity $P_N$:

$$C_P = H + P_N \Delta \quad (15)$$

It adds path diversity, but opportunity alone need not imply rerouting need.

The minimal bilinear interaction vanishes if either component is minimal and strengthens when both rise:

$$\gamma = P_N L_N, \quad 0 \leq \gamma \leq 1 \quad (16)$$

**Topology-Aware Cost (P×L):** conditions Δ on opportunity $P_N$ and need $L_N$:

$$C_{TA} = H + P_N L_N \Delta \quad (17)$$

### *D. Cost Bounds and Interpretation*

$$C_{TA} = (1 - \gamma)H + \gamma M, \quad H \leq C_{TA} \leq M \quad (18)$$

Since M − H = Δ, (17) becomes $C_{TA}$ = H + γΔ and is equivalent to (18). Values γ = 0 and 1 recover Delay and Max-Risk; intermediate values interpolate by topology. If V ≤ H, Δ = 0 and $C_{TA}$ = H.

The progression is {Delay, Backlog} → Max-Risk → Diversity → Topology-Aware: single-resource views, joint risk, opportunity, then need. L-only, $C_L$ = H + $L_N$Δ, is used only for topology ablation and is not one of the five primary costs.

## V. EXPERIMENTAL SETUP

The OMNeT++ simulator [12] with INET [13] evaluates Polska (12 nodes/18 links), NSFNET (14/21), COST266 (37/57), and Germany50 (50/88). Following the Igen-generated Australia setup of [3], the 100–1000-node topologies use 100-node increments. For centralized SDN, the controller is placed at the graph 1-median, i.e., the node minimizing the sum of physical/NED shortest-path distances to all routers. SNDlib [14] provides the Polska, COST266, and Germany50 topology references; NSFNET is retained from the shared benchmark configuration.

Following the shared configuration in [3], bidirectional links run at 1 Gbit/s; propagation uses 200,000 km/s. Each port has eight non-preemptive SPQ queues, each with a 97 kB per-queue DNC buffer-size budget. The inherited Base queueing-delay budgets for Q0–Q7 are 0.1, 0.5, 1, 2, 4, 8, 16, and 32 ms [3]. The QLSA event threshold is 10% of link bandwidth; periodic updates, admission logic, and retries remain common across costs.

Following [3], each run offers 10,000 requests, using ten requests per 10-ms interval over 10 s with randomly selected source–destination pairs. Application parameters are uniformly sampled from the ranges in Table I. The experiments were conducted on a host equipped with an AMD Ryzen 5 3500X 6-Core Processor and 16 GB of RAM.

TABLE I. FLOW REQUEST PROFILES

| Application | Deadline (ms) | Burst (bytes) | Rate (kbit/s) |
|---|---|---|---|
| Database | 80–120 | 100–400 | 300–550 |
| SCADA | 150–200 | 100–400 | 150–550 |
| Production | 10–20 | 100–400 | 100–500 |
| Control | 10–20 | 80–120 | 1–100 |
| Video | 50–200 | 140–1800 | 4000–16000 |

Small-topology cost comparisons use ten paired runs, with all five costs using the same generated request trace and random seeds within each run. Scale experiments use three runs per topology. Error bars are 95% Student-t confidence intervals (CIs) across run-level summaries; ablation intervals apply to paired differences. Architecture samples are separate, not identical-trace pairs.

Accepted flows count successfully embedded requests. Network utilization summarizes aggregate reserved bandwidth relative to data-plane capacity. Mechanism metrics average hop counts and NC post-admission delay/backlog indicators over accepted flows. Architecture overhead is reported as control packets per data-plane link per run and protocol-state memory per physical node, excluding the common best-effort baseline.

# VI. EVALUATION

## A. *Architecture Comparison*

In Table II, SDN-Dijkstra and LS-Dijkstra differ by 0.5–4.6% in mean acceptance and within 0.6 percentage points in utilization. Centralization cuts control packets per link by 45.7–86.0% and per-node protocol-state memory by 64.0–92.0%, most strongly on COST266 and Germany50. Thus, this architecture baseline shows that centralization mainly changes state management and dissemination, rather than directly causing the later routing-cost admission gains; Sec. VI-E shows the scale tradeoff.

TABLE II. ARCHITECTURE COMPARISON — MEAN ± 95% STUDENT-T CI, N = 10 PER ARCHITECTURE

| Topology | Architecture | Accepted flows | Network utilization (%) | Control packets per data-plane link | Protocol-state memory per physical node (MB) |
|---|---|---|---|---|---|
| Polska | LS-Dijkstra | 2854.1 ± 42.9 | 39.55 ± 0.46 | 2347.0 ± 122.7 | 0.2564 ± 0.0010 |
| Polska | SDN-Dijkstra | 2984.0 ± 62.3 | 39.57 ± 0.62 | 1267.9 ± 15.8 | 0.0923 ± 0.0012 |
| NSFNET | LS-Dijkstra | 2810.7 ± 32.1 | 40.40 ± 0.52 | 1863.4 ± 42.3 | 0.2759 ± 0.0007 |
| NSFNET | SDN-Dijkstra | 2878.0 ± 50.4 | 39.93 ± 0.40 | 1011.2 ± 8.6 | 0.0812 ± 0.0010 |
| COST266 | LS-Dijkstra | 4377.2 ± 62.9 | 32.71 ± 0.26 | 2885.9 ± 44.8 | 0.6837 ± 0.0006 |
| COST266 | SDN-Dijkstra | 4399.2 ± 68.1 | 32.14 ± 0.30 | 643.5 ± 3.7 | 0.0761 ± 0.0006 |
| Germany50 | LS-Dijkstra | 5951.3 ± 57.4 | 30.81 ± 0.27 | 4250.1 ± 53.2 | 1.1148 ± 0.0006 |
| Germany50 | SDN-Dijkstra | 6005.0 ± 56.8 | 30.44 ± 0.26 | 596.2 ± 6.2 | 0.0888 ± 0.0006 |

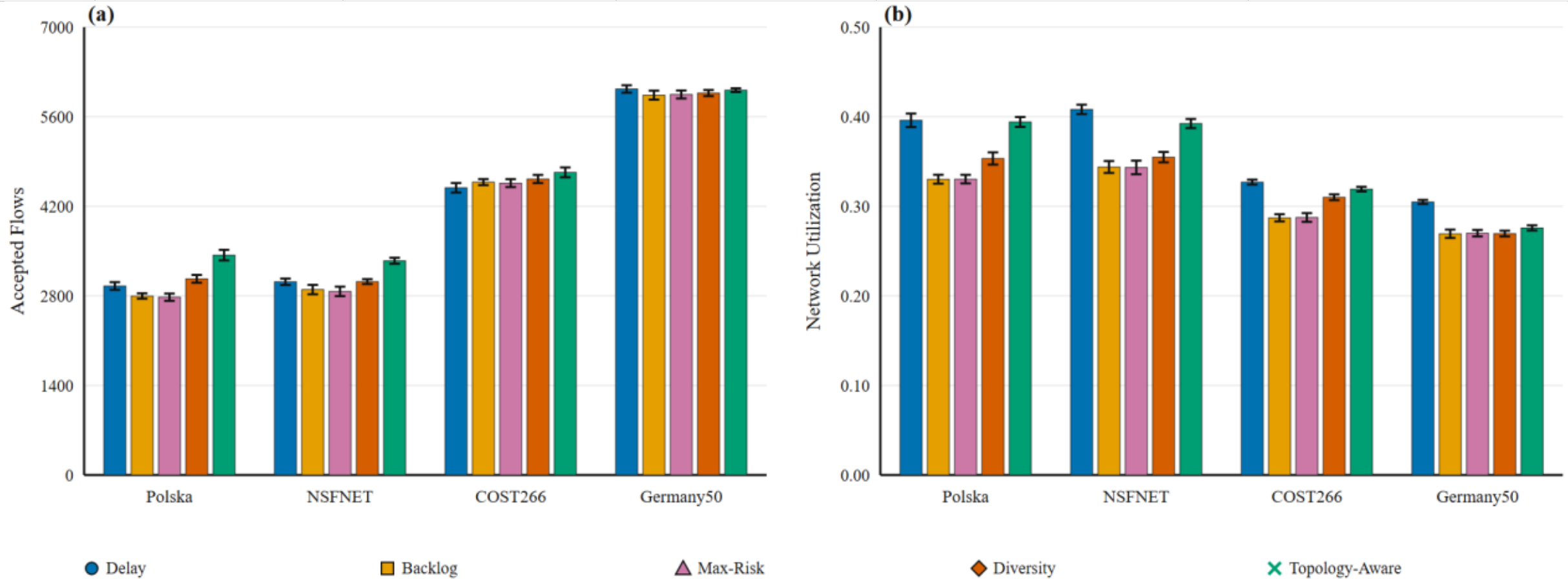

Fig. 2. Small-topology performance for five primary costs: (a) accepted flows and (b) network utilization. Error bars: 95% Student-t CIs.

## B. *Small-Topology Performance*

Fig. 2 shows the highest sample-mean acceptance for Topology-Aware on Polska, NSFNET, and COST266; Germany50 is the exception, with Delay slightly higher. Gains over Max-Risk are strongest on Polska and NSFNET, smaller on COST266, and weak and uncertain on Germany50.

The admission gains are not accompanied by a uniform increase in aggregate utilization, indicating that aggregate bandwidth use alone does not explain acceptance. This is consistent with topology-conditioned ranking preserving useful feasible queue resources for later requests.

## C. *Topology Ablation*

Fig. 3 tests whether either topology descriptor is sufficient in isolation. Positive values denote more accepted flows than Max-Risk, while a paired confidence interval crossing zero indicates that the observed difference is inconclusive. P-only and L-only effects are topology dependent: L-only is inconclusive on NSFNET, and both single-descriptor intervals cross zero on COST266. In contrast, P×L gives the largest mean gain over Max-Risk on all four topologies and a clearly positive paired gain on COST266. Germany50 remains weak and uncertain because the relevant intervals cross zero.

P represents rerouting opportunity, whereas L represents rerouting need. High P alone does not imply rerouting is needed, and high L alone does not guarantee useful alternatives. The ablation therefore supports conditioning queue risk on both structural dimensions rather than relying on either descriptor alone. P×L is an interaction, not an additive decomposition.

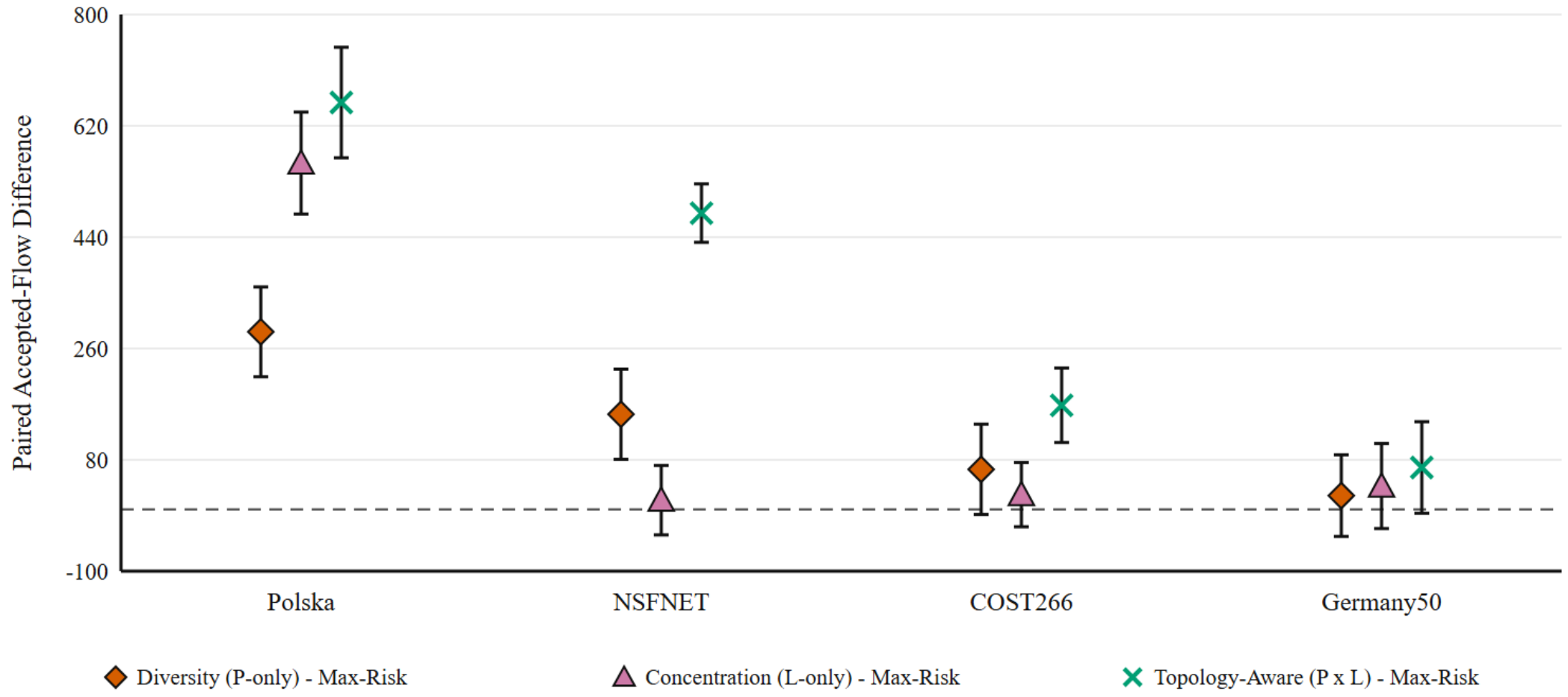

Fig. 3. Paired accepted-flow differences relative to Max-Risk. P-only is Diversity; L-only is used only for ablation. Error bars: 95% Student-t CIs.

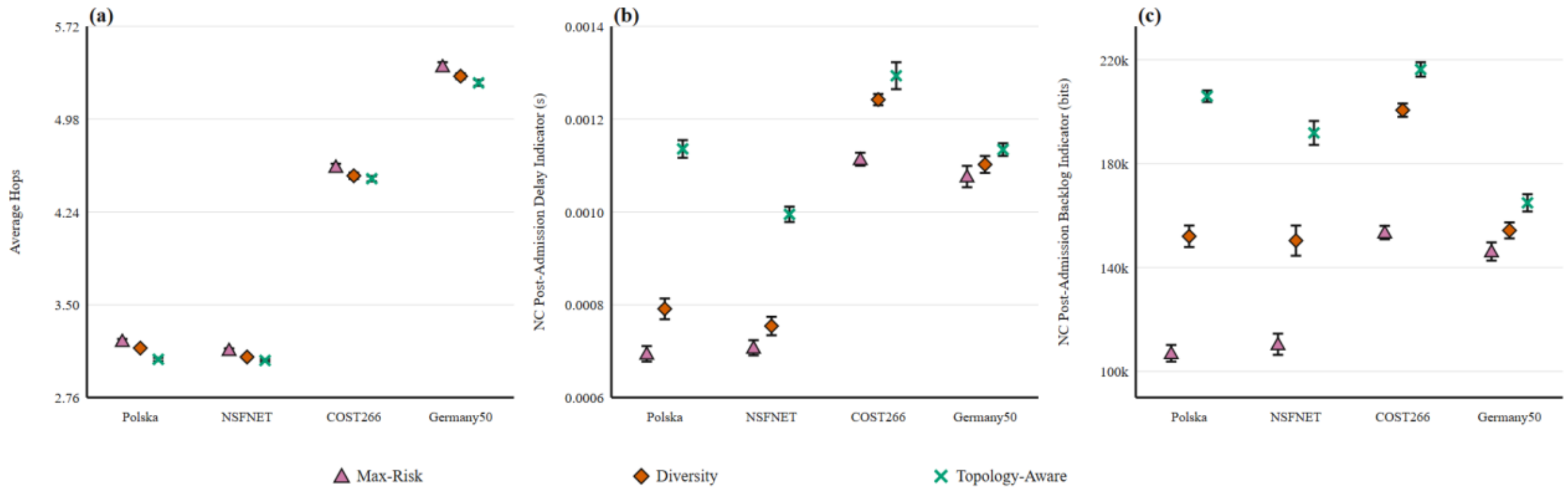

Fig. 4. Mechanism analysis over accepted flows: (a) average hops, (b) NC post-admission delay indicator (s), and (c) NC post-admission backlog indicator (bits). Error bars: 95% Student-t CIs.

### D. *Mechanism Analysis*

Fig. 4 separates path-length effects from queue-resource use. Average hops measures mean path length over accepted flows, while the NC delay and backlog indicators are the corresponding post-admission analytical queueing-delay and backlog bounds averaged over accepted flows. Larger indicators that remain within the admission limits denote greater use of the feasible delay and buffer envelope, not constraint violations. In Fig. 4(a), mean hop count decreases from Max-Risk to Diversity and again to Topology-Aware on all four topologies, so the admission gain is not achieved through longer-path detours. Figs. 4(b) and 4(c) show larger feasible NC bounds among accepted flows, consistent with topology-conditioned costs using more of the queue-resource envelope.

The pattern is strongest on Polska and NSFNET, where admission gains are also largest, and weakest on Germany50. This correspondence is descriptive rather than causal because the accepted-flow populations differ across methods.

### E. *Large-Scale Evaluation*

In Fig. 5(a), Topology-Aware exceeds Delay by about 15.4% at 100 nodes and 26.3% at 1000 nodes. Across 100–1000 nodes, Delay and the Dijkstra baselines decline by about 9.8–11.3%, versus only 1.3–3.9% for the risk-aware costs. The practical separation between delay-only and risk-aware routing therefore widens with topology size in this scale set. Advanced risk-aware variants remain tightly grouped, with a mean spread of only about 0.4–2.0% at each scale point, suggesting that the main large-scale benefit in this scale set is associated with incorporating backlog risk rather than with topology conditioning alone.

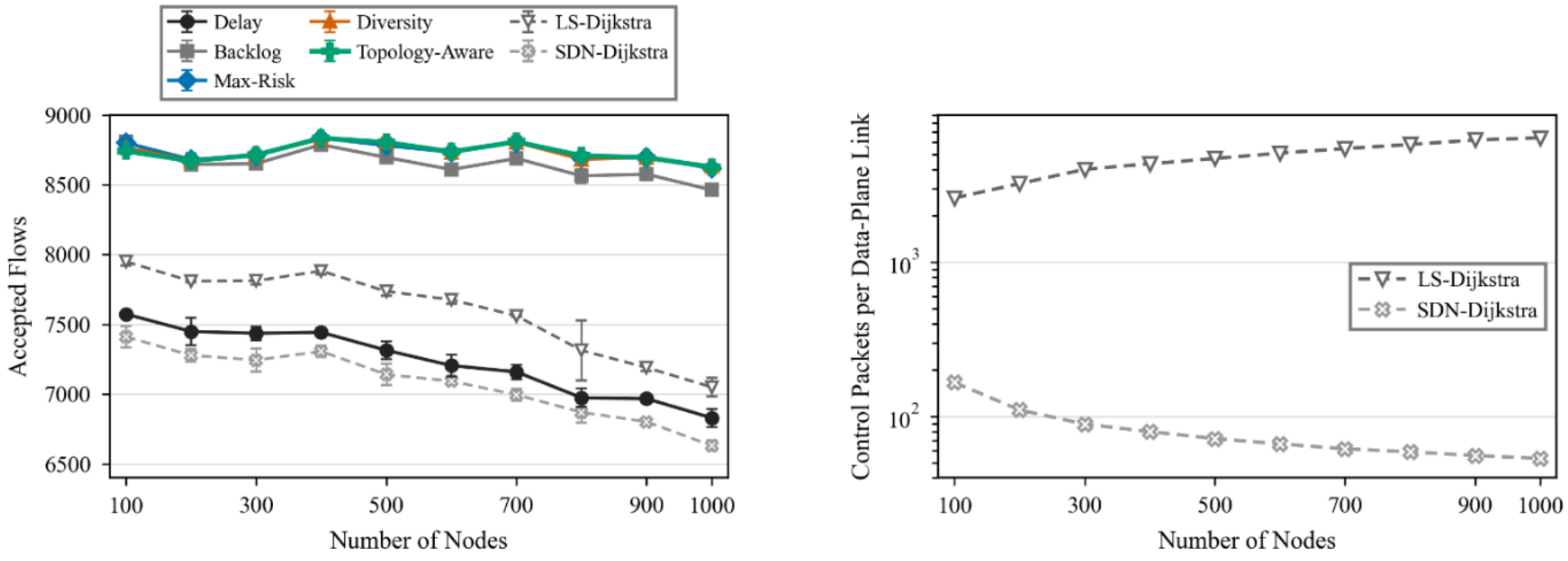

(a) Accepted flows

(b) Control overhead

Fig. 5. Large-scale evaluation on 100–1000-node topologies: (a) accepted flows for the five routing costs and the LS-Dijkstra and SDN-Dijkstra architecture baselines; (b) control packets per data-plane link for the two architecture baselines. Accepted-flow error bars show 95% Student-t CIs.

LS-Dijkstra admits more flows than SDN-Dijkstra throughout the scale experiment. Control packets per data-plane link measure control-plane communication burden normalized by the number of data-plane links; Fig. 5(b) uses a logarithmic y-axis. LS-Dijkstra overhead rises with scale, whereas SDN-Dijkstra remains much lower and decreases. Under the fixed 10,000-request workload, centralized reporting avoids network-wide flooding and spreads control activity over a larger data plane; LS-Dijkstra disseminates queue-state updates across the growing topology.

These baselines expose a performance–overhead tradeoff. Routing-cost and control-plane scalability are separate: centralization does not impose a uniform admission gain or penalty, but its overhead advantage is more consistent than its routing-performance effect. SDN-Dijkstra versus the LARAC variants is not a pure Dijkstra-versus-LARAC ablation because routing objectives and path selection also differ.

## VII. DISCUSSION AND CONCLUSION

We evaluate a five-cost progression from single-resource Delay and Backlog to joint Max-Risk, Diversity (P-only), and Topology-Aware (P×L), where P denotes rerouting opportunity and L rerouting need.

Topology-Aware has the highest mean among five costs on three small topologies and exceeds Max-Risk in mean on all four, although Germany50 remains weak and uncertain. On COST266, neither single-descriptor ablation is conclusive, whereas joint P×L conditioning yields a clear positive gain. From Max-Risk to Diversity to Topology-Aware, admission rises, hop count falls, and feasible NC indicators increase; this descriptive pattern is consistent with broader feasible queue-state use rather than longer paths or higher aggregate utilization.

At scale, the delay-only versus risk-aware gap widens while advanced risk-aware variants remain close. Centralized state management consistently reduces dissemination overhead but does not ensure higher admission, so architecture and cost effects remain separate. Empirical P/L normalization ranges may require recalibration for different topology families. Overall, static topology structure provides useful context for deciding how strongly post-admission backlog risk should influence queue-level routing.